\documentclass[10pt]{NSP1}
\usepackage{url,floatflt}
\usepackage{helvet,times}
\usepackage{psfig,graphics}
\usepackage{mathptmx,amsmath,amssymb,bm}
\usepackage{float}
\usepackage[bf,hypcap]{caption}
\usepackage{algorithmic}
\usepackage{algorithm}
\usepackage{array}

\emergencystretch=\maxdimen
\def\no{\noindent}

\def\firstpage{107}
\def\thevol{9}
\def\thenumber{2}
\def\theyear{2020}
\begin{document}

\titlefigurecaption{{\large \bf \rm Information Sciences Letters }\\ {\it\small An International Journal}}

\title{IoT and Neural Network-Based Water Pumping Control System For Smart Irrigation}

\author{M. E. Karar\hyperlink{author1}{$^{1,2,\star}$}, M. F. Al-Rasheed\hyperlink{author2}{$^{1}$}, A. F. Al-Rasheed\hyperlink{author3}{$^{1}$} and Omar Reyad\hyperlink{author4}{$^{1,3}$}}
\institute{$^1$College of Computing and Information Technology, Shaqra University, Shaqra, KSA \\
           $^2$Faculty of Electronic Engineering, Menoufia University, Menoufia, Egypt\\
           $^3$Faculty of Science, Sohag University, Sohag, Egypt}

\titlerunning{IoT and Neural Network-Based Smart Irrigation}
\authorrunning{M. E. Karar et al.}

\mail{mkarar@su.edu.sa}

\received{2 Feb. 2020}
\revised{13 April 2020}
\accepted{17 April 2020}
\published{1 May 2020}

\abstracttext{This article aims at saving the wasted water in the process of irrigation using the Internet of Things (IoT) based on a set of sensors and Multi-Layer Perceptron (MLP) neural network. The developed system handles the sensor data using the Arduino board to control the water pump automatically. The sensors measure the environmental factors; namely temperature, humidity, and soil moisture to estimate the required time for the operation of water irrigation. The water pump control system consists of software and hardware tools such as Arduino Remote XY interface and electronic sensors in the framework of IoT technology. The machine learning algorithm such as the MLP neural network plays an important role to support the decision of automatic control of IoT-based irrigation system, managing the water consumption effectively.}


\maketitle

\section{Introduction}  \label{intro}
Water irrigation is the essential element for the agricultural sector \cite{c0}. Therefore, the farmers need an advanced technology to manage their corps with appropriate levels of water irrigation. According to the recent development of science and technology \cite{c01}, the Internet of Things (IoT) can solve the above agricultural problem by automatically controlling the water pump using machine learning algorithms and embedded systems technology. Measuring the environmental factors such as temperature and soil moisture via a set of sensors will support the artificial neural networks controller-based Arduino board to handle the rates of water irrigation level in an efficient manner \cite{c1}. The developed irrigation control system includes software and hardware tools such as Arduino Uno microcontroller board and Remote XY for cloud Internet connection with Android user interface (UI) \cite{c2}. Machine learning techniques integrated with the embedded system of pump controller will support the farmers to decide the required quantity of water for the operation of farm irrigation precisely. One of the problems facing the world is water consumption and finding techniques and technologies to save it \cite{c3}. Moreover, the remote water irrigation management can be handled during the Coronavirus pandemic and the period of quarantine \cite{c4}. 

The smart irrigation management plays an important role to increase crop and to decrease costs with contributing to the climate adaptation and sustainability \cite{c5}. The IoT presents the core of the smart irrigation systems, including the internetwork of physical devices, embedded electronic boards, software, sensors, and communication among these items to exchange data over the Internet connection \cite{c6}. Therefore, the IoT-based powered irrigation system can save the water use with improving the productivity and quality of the crops. This study is focused on exploiting the IoT capabilities to control water irrigation precisely.  
  
\begin{figure*}[h]
\centering
\includegraphics[width= 0.60\textwidth]{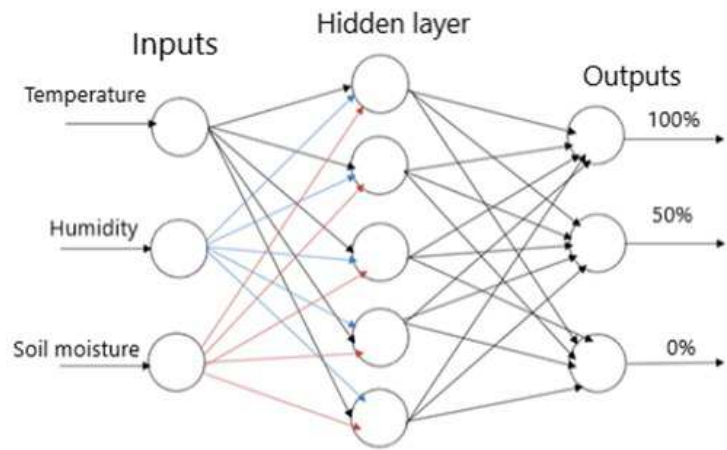}
\caption{The proposed MLP neural network controller for embedded water irrigation system}
\label{Fig.1}       
\end{figure*}

\begin{figure*}[h]
\centering
\includegraphics[width= 0.65\textwidth]{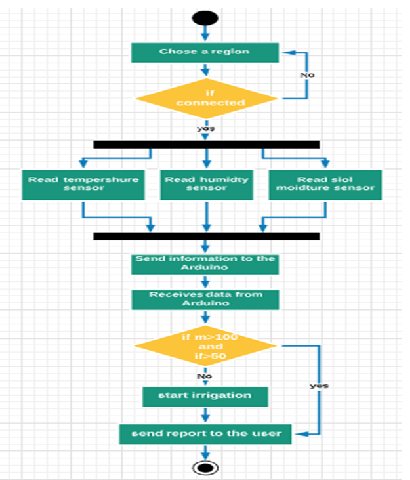}
\caption{Activity diagram of the proposed water irrigation control system}
\label{Fig.2}       
\end{figure*}

In the previous studies, subsurface micro irrigation techniques, e.g. capillary irrigation, have been used to reduce water use for cultivation \cite{c7}. They provide water transfer directly to the rooting zone of plant. Based on measuring the water content in the soil, Abidin et al. \cite{c7} proposed an intelligent irrigation control system to the field at right amounts and times. However, other important environmental factors such as humidity were ignored. On the other side, the technology of wireless sensor networks (WSN) has been applied to the agriculture sector to offer new trends like Precision Agriculture (PA) \cite{c8}. Emerging the WSN with advanced knowledge-based fuzzy controller \cite{c9} were proposed to automate the water levels of irrigation system \cite{c10}. Moreover, using embedded and electronic boards such as STC89C52 MCU (Micro Control Unit) was proposed to provide a controller for smart irrigation system \cite{c11}. 

Artificial neural networks such as multilayer perceptron (MLP) and radial basis function (RBF) neural networks, and deep neural networks have been widely used in many applications \cite{c12,c13,c14}. These neural networks showed powerful performance to deal with real problems in an efficient way by initial training of labeled data, and then autonomously operated \cite{c15,c16}. This study uses the feedforward MLP neural networks as a smart controller for the proposed smart water irrigation system \cite{c17}.

Considering the Vision 2030 of Saudi Arabia, the utilities of IoT technology with artificial intelligence will be applied to achieve remote farm irrigation management and saving wasted water consumption in the agricultural field. Therefore, this article aims at developing a new smart irrigation system to control the water pump using the Internet of Things (IoT) and MLP neural networks.

The rest of the paper is organized as follows. In Section \ref{Sec1}, materials and the proposed smart irrigation system are presented. Hardware implementation and results are described in Section \ref{Sec2} while conclusions and the future work are depicted in Section \ref{Sec3}.

\section{Materials and Methods} \label{Sec1}
\subsection{Materials} \label{Sec1.1}
In this study, main components of the embedded system have been used such as the environmental sensors and Wi-Fi unit (ESP8266) that receives data from the cloud and sends it to Arduino UNO, Arduino UNO controls the opening and closing of DC Motors and the water is pumped through the plastic water tubes.

Arduino IDE was used to program sensors on Arduino UNO. Remote XY app was used to provide an electronic cloud and display data to the user. The MATLAB was used to design and train MLP neural network, producing the required values of trained neural weights.

\subsection{Proposed Smart Irrigation System} \label{Sec1.2}
Figure \ref{Fig.1} shows the schematic diagram of MLP neural network as a water irrigation controller includes three layers. The first layer is the input layer, which includes three inputs from the sensor data (temperature, humidity, and soil moisture). The second layer is the hidden layer includes 5 nodes to support the final decision of irrigation system. Finally, the third layer is the output layer, including three options for controlling the water irrigation system based the percentage operation of water pumping as 100\% (full operation), 50\% (half operation), and 0\% (water pump is off). Table \ref{Tab.1} illustrates the simplified IF-Then rule-base of controlling the water pump, based on the expected range of humidity, temperature, and soil moisture \cite{c18}.

\begin{figure*}[h]
\centering
\includegraphics[width= 0.70\textwidth]{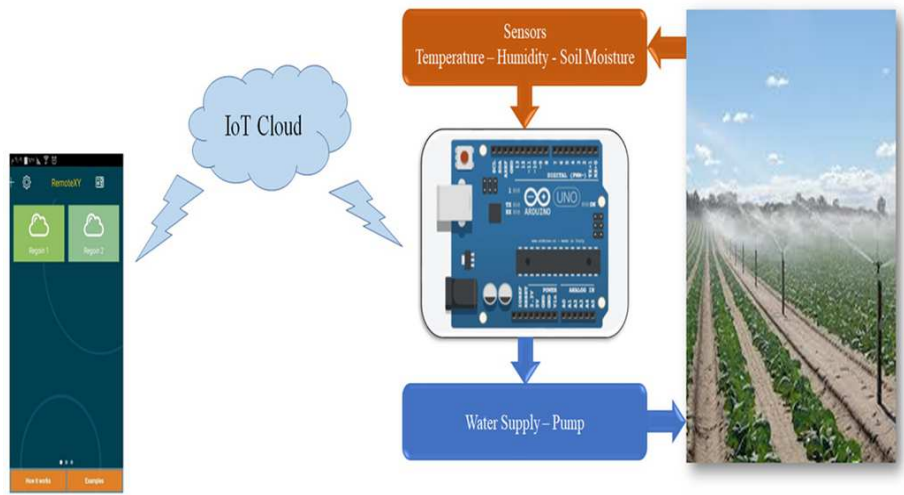}
\caption{Proposed IoT-based system design of smart water irrigation including Arduino Board, environmental sensors, and a water pump.}
\label{Fig.3}       
\end{figure*}

Figure \ref{Fig.2} shows the graphical activity diagram to demonstrate how the program works, as described above. The program starts and the field is searched. If the range is found, the specified range is entered, and the sensor data is fetched so that the water pumping amount is set from $100$\% to $50$\% until the pump closes. Figure \ref{Fig.3} depicts the main hardware components and connections of the developed control system design, as described above in Materials section.

\begin{figure*}[h]
\centering
\includegraphics[width= 0.70\textwidth]{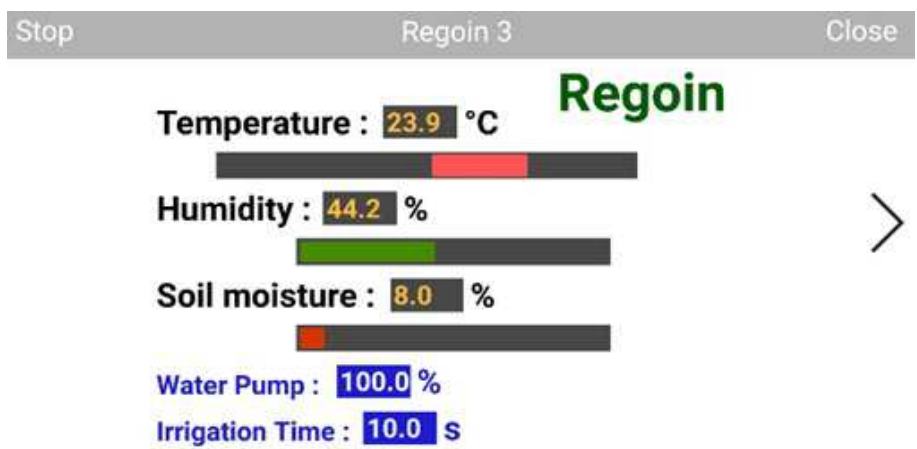}
\caption{Mobile User Interface (UI) of the developed irrigation system with a case of full water irrigation pumping (100\%) in 10 seconds.}
\label{Fig.4}       
\end{figure*}

\section{Hardware Implementation and Results} \label{Sec2}
The user Interface (UI) using RemoteXY Arduino has been developed to display the sensor data, namely the temperature sensor, moisture sensor, and soil moisture sensor. According to the simplified rules in Table \ref{Tab.1}, the amount of water irrigation is determined and pumped in the farm with the expected operational time, as depicted in Figure \ref{Fig.4}.

\begin{figure*}[h]
\centering
\includegraphics[width= 0.65\textwidth]{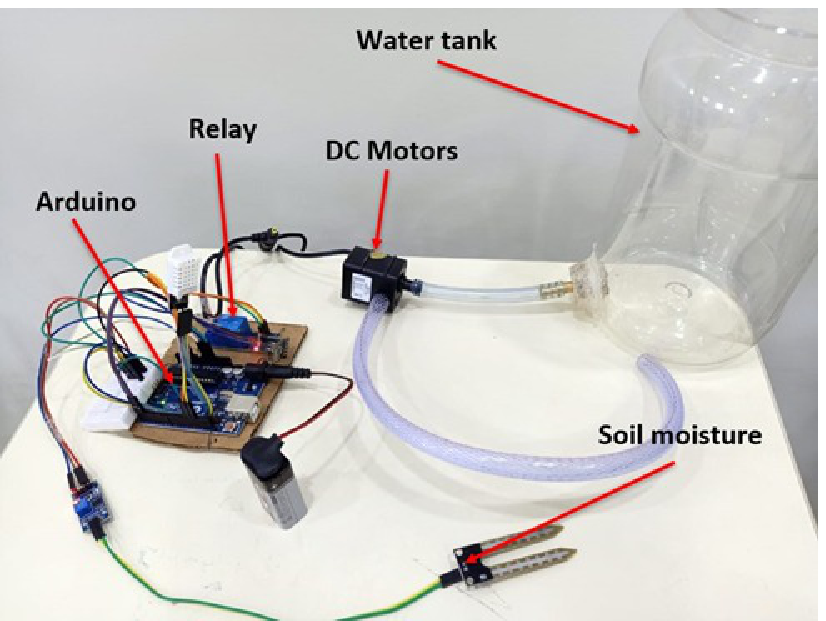}
\caption{Experimental setup of embedded water irrigation control system.}
\label{Fig.5}       
\end{figure*}

In Figure \ref{Fig.5}, the hardware implementation of water irrigation control system is depicted including the Arduino Uno board and a $12$ DC volt water pump with external relay. The remote XY is linked via Wi-Fi module and the IoT cloud to give full information about sensor data and the operational status of water pump.

\section{Conclusion} \label{Sec3}
This project presented a new design of water pump control for the development of smart irrigation system linked with a mobile application. The developed irrigation system supports the farmers to save wasted water, time and effort to increase the productivity of their crops. 

In the future work of this project, case studies of different crops will be used, and also automatic measuring the level of the water the tank to alert the farmers via secured Short Message Service (SMS) mobile messages \cite{c19,c20,c21}. Finally, applying the developed automatic irrigation system in the farms of Shaqra city, Saudi Arabia.

{\begin{table*}[h]
\caption{Simplified rule-base of controlling water pump based on the values of humidity, temperature, and soil moisture.}
\label{Tab.1}
\begin{center}
\begin{tabular}{| c | c | c | c |} \hline 
  \multicolumn{3}{|c|}{\textbf{Inputs}} & \textbf{Output}                   \\ \hline
\textbf{Temperature} & \textbf{Humidity} & \textbf{Soil moisture} & \textbf{Water pump operation } \\ 
($^\circ$C) & (\%)  & (\%)   & (\% of full load) \\ \hline
Low     & Low     & Low      & Water pump is ON    \\ 
($<$ 25)  &  ($<$ 40) &  ($<$ 10)  &   (100.0 \%)   \\ \hline
Low     & Low     & Medium   & Water pump is ON    \\ 
($<$ 25)  & ($<$ 40)  & (10-20)  &  (50.0 \%)   \\ \hline
Medium  & Medium  & Medium   & Water pump is ON    \\ 
(25-35) & (40-70) & (10-20)  & (50.0 \%) \\ \hline
High    & High    & High     & Water pump is Off   \\ 
($>$ 35) or  & ($>$ 70) or  &  ($>$ 20) or  & (0.0 \%)  \\ 
other conditions    & other conditions    & other conditions     &    \\ \hline
\end{tabular} 
\end{center}
\end{table*}} 

\section*{Acknowledgment}
This project is financially supported by the College of Computing and Information Technology, Shaqra University.

\emergencystretch=\hsize

\begin{center}
\rule{6 cm}{0.02 cm}
\end{center}

\end{document}